\documentclass[prd,10pt,preprintnumbers,twocolumn,numbers,sort&compress,nofootinbib,floatfix]{revtex4}
\usepackage{amssymb}  
\usepackage{amsfonts} 
\usepackage{amsmath}
\usepackage[hypertex]{hyperref}
\usepackage{psfrag}
\usepackage{graphicx}

\newtheorem{rule-def}[theorem]{Rule}

\newcommand{\be}{\begin{equation}}
\newcommand{\ee}{\end{equation}}
\newcommand{\bea}{\begin{eqnarray}}
\newcommand{\eea}{\end{eqnarray}}

\begin{document}
\date{\today}

\title{SEARCHING FOR A SOLUTION TO THE AGE PROBLEM
OF THE UNIVERSE}

\author{Saibal Ray\dag\footnote{e-mail:saibal@iucaa.ernet.in (Corresponding Author)}
 and Utpal Mukhopadhyay\ddag}

\address{\dag\ Department of Physics, Barasat Government College, Kolkata
700 124, North 24 Parganas, West Bengal, India and IUCAA, Post Bag
4, Pune 411 007, India}

\address{\ddag\ Satya Bharati Vidyapith,
Kolkata 700 126, North 24 Parganas, West Bengal, India}

\begin{abstract}
We present  here a  phenomenological cosmological model under
perfect fluid distribution with a stiff equation of state
$p=\rho$. The erstwhile cosmological constant is assumed to be a
time dependent variable, i.e., $\Lambda =  \Lambda(t)$ in our
study. It has been shown that the estimates of different
cosmological parameters from this model are in good agreement with
the experimental results, especially $13.79$ Gyr as the age of the
universe is quite satisfactory. The behavior and relation of
$\Lambda$-stiff fluid model with dust, viscous fluid and variable
$G$ have also been investigated in detail.
\end{abstract}

\maketitle


``The problem of the age of the observable universe - the time
back to the `big bang' - has puzzled laymen, churchmen and
scientists for many years.'' - W. A. Fowler (in {\it The Age of
the Observable Universe})~\cite{fow86}\\

{\bf 1.~Introduction}\\

The cosmological picture that emerges after the observations of
SNe Ia by HZT and SCP teams reveals that at present we are
residing in an accelerating universe \cite{perl98,riess98} whose
geometry is Euclidean in nature \cite{sievers03}. This speeding up
of the universe started about $7$ Gyr ago \cite{kirs03} and some
kind of repelling force, termed as {\it dark energy}, is supposed
to be responsible for catapulting the once decelerating Universe
into an accelerating one. Using various observational data it has
also been possible to pin down, to some extent, the ranges of
various parameters of the Universe at the present era. For
instance, the acceptable limit of $H_0$, the Hubble parameter at
present, is $72 \pm 8$ Km~s$^{-1}$Mpc$^{-1}$ \cite{alta04} and
while taking into consideration of various ranges as proposed by
different workers from varying standpoints \cite{ferrer01} the
currently accepted observational value of the age of the universe
becomes $14 \pm 0.5$ Gyr \cite{chab98} (an elaborated data sheet
has been produced in ref. \cite{ray04}).

Many variants of $\Lambda$, a dark energy representative, have
been proposed by different workers to account the cosmological
consequences. Since, supernovae Ia observations predict a small
value of $\Lambda$ at the present epoch \cite{podar00}, now-a-days
$\Lambda$ with dynamical character is preferred  over a constant
$\Lambda$~\cite{over98}. This has been assumed to vary with time
so that decrease of $\Lambda$ from a large initial value to its
present small value can easily be realized.

Mainly, in the theoretical point of view, the possibility of a
nonzero and varying $\Lambda$ came into picture in connection to
the {\it age problem} of the universe. This is because of the fact
that for large $\Lambda$ the age of the universe can, in
principle, become infinite~\cite{wein72,over98}. Even with the
time-decaying $\Lambda$, it is seen that the present age of the
universe either suffers from low-age problem \cite{over98} or it
is as large as $27.4 \pm 5.6$ Gyr \cite{viswa02}. In connection to
the low age we can mention the results of Overduin and
Cooperstock~\cite{over98} where for the Hubble parameter $H_0=73
\pm 10$ kms$^{-1}$Mpc$^{-1}$ the age of the universe in its lower
limit is $5.4$ Gyr! The situation improves slightly for lower
value of $H_0(=55 \pm 10$ kms$^{-1}$Mpc$^{-1}$)~\cite{sand96} when
it becomes $7.6$ Gyr only. This values are much lower than the
estimated globular cluster ages which are thought to be in the
range $9.6$ Gyr~\cite{chab98} to $13 \pm 3$ Gyr~\cite{ccq86} and
very recent observations claim that it is most probably $12.5 \pm
1.2$ Gyr~\cite{glr01,cay01}. However, in the standard model of a
flat $\Lambda$-dominated universe the age of the universe is $13.7
\pm 0.2$ Gyr as obtained by Spergel et al.~\cite{sper03} whereas
Kunz et al.~\cite{kunz04} in the $\Lambda$-CDM case find out value
as $13.55 \pm 0.26$. For a slightly closed $\Lambda$-CDM universe
with $H_0 = 66^{+6.7}_{-6.4}$ kms$^{-1}$Mpc$^{-1}$ Tegmark et
al.~\cite{teg04} estimate an age of $14.1^{+1.0}_{-0.9}$ Gyr.

Under this background, in the present article we have investigated
three types of widely used \cite{al96, viswa01} phenomenological
forms of kinematical $\Lambda$, viz., $\Lambda \sim  (\dot
a/a)^2$, $\Lambda \sim  (\ddot a/a)$ and $\Lambda\sim \rho$, where
$a$ is the scale factor of the Robertson-Walker metric and $\rho$
is the energy density of the universe (a detail account of these
phenomenological $\Lambda$ models is available in ref.
\cite{ray04}).

As a result of the present investigations under the different
phenomenological models as mentioned above we have found out the
equivalence of the $\Lambda$-models through the parameters
$\alpha$, $\beta$ and $\gamma$ involving in the models. Under the
perfect fluid distribution, specially with a stiff equation of
state $p=\rho$, the age and the other physical parameters have
been estimated. Surprisingly, the age $13.79$ Gyr is in good
agreement with the present available experimental results. The
behavior and relation of stiff fluid to dust and viscous fluid
also have been critically investigated. It is also observed that
Variable $\Lambda$ and variable $G$ models show some unique common
features in connection to the present accelerating universe.

The article has been organized as follows: in Section 2 the
standard general relativistic Einstein field equations are
presented with the inclusion of time varying cosmological
constant, viz., $\Lambda =  \Lambda(t)$, in the energy-momentum
tensors. Section 3 deals with the solutions of various
phenomenological models within this framework of modified general
relativity. In the Section 4 some salient features of the present
model are discussed. Section 5 presents concluding remarks on the
status of the stiff fluid and also the problems related to
the age of the universe.\\

{\bf 2.~Einstein Field Equations}\\

Let us consider the Robertson-Walker metric which is given by \bea
ds^2  =  -dt^2 +a(t)^2\left[\frac{dr^2}{1  -  kr^2} +  r^2
(d\theta^2 + sin^2\theta d\phi^2)\right] \eea  where $k$ is the
curvature constant which takes the specific values $-1$, $0$ and
$+1$ respectively for open, flat and close models of the universe
and $a$ is the cosmic scale factor as mentioned earlier.

With the assumption that the so-called {\it cosmological constant}
is time dependent here, viz., $\Lambda = \Lambda(t)$ and $c$, the
velocity of light in vacuum  is assumed to be unity in
relativistic units, the Einstein field equations \bea R^{ij}
-\frac{1}{2}Rg^{ij}   =  -8\pi   G\left[T^{ij}   -
\frac{\Lambda}{8\pi G}g^{ij}\right] \eea for the above spherically
symmetric metric (1) yield, respectively, the Friedmann and
Raychawdhuri equations \bea \left(\frac{\dot a}{a}\right)^2 +
\frac{k}{a^2} = \frac{8\pi G}{3}\rho +\frac{\Lambda}{3}, \eea \bea
\frac{\ddot a}{a} =  - \frac{4\pi G}{3} (\rho +  3p)+
\frac{\Lambda}{3} \eea where $G$, $\rho$ and $p$ are,
respectively, the gravitational constant, matter-energy density
and fluid pressure.

Also, the energy conservation law can be written as \bea 8\pi G(p
+ \rho)\frac{\dot a}{a} =-\frac{8\pi G}{3}\dot \rho -
\frac{\dot\Lambda}{3}. \eea
 \noindent
Let us now choose the barotropic equation of state in the form
\bea p = w\rho, \eea where $w$ is a constant, known as the
equation of state parameter and can take the values $0$, $1/3$ and
$1$ respectively for the pressureless dust, electromagnetic
radiation and stiff or Zel'dovich fluid.

Using this barotropic equation of state (6), in a straight forward
way, the equation (4) can be written as \bea \frac{\ddot a}{a}  +
 \frac{4\pi  G}{3}  (1   +  3w)\rho  =\frac{\Lambda}{3}.
\eea After performing differentiation of equation (3) with respect
to time and using equations (4) - (7) for elimination of $\rho$,
one obtains the following equation \bea \left(\frac{\dot
a}{a}\right)^2   +
 \left[3\left(\frac{1+w}{1 + 3w}\right) - 1 \right]\frac{\ddot a}{a} +
 \frac{k}{a^2} =\left(\frac{1+ w}{1 + 3w}\right)\Lambda.
\eea This is the general equation for our investigation in the
presence of the curvature parameter $k$.\\

{\bf 3.~Solutions}\\

 At this point let us consider the {\it ansatz} $\Lambda = 3\alpha(\dot a/a)^2$,
 where   $\alpha$  is   a  constant.  Now, inflation theory predicts
 and CMB detectors such as BOOMERanG \cite{Bernardis00,Bernardis01,Netterfield02},
 MAXIMA \cite{Hanany00,Lee01,Balbi01}, DASI \cite{Halverson02}, CBI
\cite{Sievers03} and WMAP \cite{Bennett03,Spergel03} confirm that
the universe is spatially flat. Therefore, for the flat universe
where $k = 0$, the equation (8) reduces to \bea 2a \ddot a + (1 +
3w  - 3w\alpha -3\alpha)\dot a^2 = 0. \eea

  The general solutions of the above equation (9)
can be given as
\bea a(t) = C_1t^{2/3(1 - \alpha)(1 + w)}, \eea
\bea \rho(t) = \frac{1}{6\pi G(1 - \alpha)(1 + w)^2}t^{-2}, \eea
\bea \Lambda(t) = \frac{4\alpha}{3(1 - \alpha)^2(1 + w)^2}t^{-2}
\eea
where  $C_1$ is  an integration constant and $0 < \alpha < 1$
for physical validity.

Therefore, for stiff fluid distribution ($w=1$) the proportional
relation of the cosmic scale factor to the cosmic age becomes $
t^{1/3(1-\alpha)}$. In a similar way, it can be shown that for the
{\it ansatz} $\Lambda  = \beta (\ddot a/a)$ and $\Lambda  = 8\pi
G\gamma  \rho$ the cosmic scale  factor $a(t)$ is, respectively,
proportional to $t^{(\beta-1)/(\beta-3)}$ and $t^{(\gamma+1)/3}$,
where $\beta$ and $\gamma$ are another two constants like
$\alpha$. For all these three models, as are evident from
equations (11) and (12), the cosmic matter-energy density
$\rho(t)$ and the cosmological parameter $\Lambda(t)$ both follow
an inverse square law with $t$.

Now, in absence of any curvature, the cosmic matter-energy density
parameter $\Omega_m(=  8\pi G  \rho/3H^2)$ and cosmic
vacuum-energy density parameter $\Omega_{\Lambda} (=\Lambda/3H^2)$
are related by \bea \Omega_m +  \Omega_{\Lambda} = 1  . \eea This
result is consistent with the current constraint on cosmic density
  parameters \cite{tarun04}. Expressing the solutions of
(8) for stiff fluid in terms of $\Omega_m$, it has been found that
for all the three $\Lambda$-models, $a(t)$ is proportional to
$t^{1/{3\Omega_m }}$ whereas the expressions for $\rho(t)$ and
$\Lambda(t)$ are identical. Moreover, $\alpha$, $\beta$ and
$\gamma$ are related to $\Omega_m$ and $\Omega_{\Lambda}$ by \bea
\alpha = \Omega_{\Lambda}, \quad \beta
=\frac{3\Omega_{\Lambda}}{\Omega_{\Lambda}-2\Omega_m},   \quad
\gamma =\frac{\Omega_{\Lambda}}{\Omega_m}. \eea Equation (14),
with the help of equation (13), enables us to have interrelation
between $\alpha$, $\beta$ and $\gamma$ as follows
 \bea \alpha  =
\frac{2\beta}{3(\beta - 1)} =\frac{\gamma}{1 + \gamma} .
\eea

This clearly shows that the three forms, viz., $\Lambda = 3\alpha
(\dot a/a)^2$, $\Lambda = \beta (\ddot a/a)$ and $\Lambda = 8\pi G
\gamma \rho$, are equivalent and the three parameters $\alpha$,
$\beta$ and $\gamma$ are interconnected by the relation (15).
Therefore, it is possible to search for the identical physical
features of others if any one of the phenomenological
$\Lambda$ model is known.\\

{\bf 4.~Physical Features}\\

{\bf 4.1.~Age of the universe from the stiff fluid model}\\
 The most  striking and puzzling feature of this
stiff fluid model is the fine agreement of the physical
parameters, especially the present age of the universe, with the
observational results. To get an explicit expression for age let
us differentiate equation (10) and divide it by the cosmic scale
factor $a$, which ultimately yields \bea t = \frac{2}{3(1 -
\alpha)(1 + w)H}. \eea In the case of stiff fluid ($w=1$) the
above equation (16), by the use of the equations (13) and (14),
reduces to \bea t= \frac{1}{3\Omega_{m}H}. \eea Recent
measurements indicate that the value of $\Omega_{m0}$, the cosmic
matter-energy density parameter at present, is $\Omega_{m0} = 0.33
\pm 0.035$ \cite{viswa02}. Assuming $\Omega_{m0}  = 0.295$, the
lowest acceptable limit of this range, and $H_0 =
72$kms$^{-1}$Mpc$^{-1}$,
 the present age of the universe, $t_0 (= 1/3\Omega_{m0}H_0)$  is  found to be
$13.79$ Gyr which is in excellent agreement with the  age of the
universe as calculated on the basis of WMAP  data \cite{kirs03} as
well as the current accepted age mentioned earlier. A comparison
between the ages for other values of $\omega$ shows a situation
which goes in favor of stiff fluid (TABLE I).

\begin{table*}
\begin{minipage}{105mm}
\caption{Age of the universe from different $\omega$ models}
\label{tab 1}
\begin{tabular}{@{}llrrrrlrlr@{}}
\hline $\Omega_{m0}$  &&$H_0$                   &&$t_0$\\
               &&(km~s$^{-1}$Mpc$^{-1}$)  &&(Gyr)\\
               &&                        &&$\omega =1$ &&$\omega =0$ &&$\omega =1/3$\\

\hline 0.295          &&64                      &&15.53 &&31.07
&&25.91 \\
               &&72                      &&13.79 &&27.61 &&22.83\\
               &&80                      &&12.43 &&24.86 &&20.70\\
\hline
\end{tabular}
\end{minipage}
\end{table*}

Moreover, the values of $\rho_0 (= 1/24 \pi G \Omega_{m0}
{t_0}^2)$  and $q_0$, respectively the present values of the
cosmic matter-energy density $\rho$ and the deceleration parameter
$q$, as calculated on the basis of $\Omega_{m0}  = 0.33 \pm 0.035$
and $H_0  = 72$kms$^{-1}$Mpc$^{-1}$, confirm that the present
models support the idea of an open as well as accelerating
universe \cite{efs98}. Also, the  value of $\Lambda_0 (= (1 -
\Omega_{m0})/3 {\Omega_{m0}}^2{t_0}^2)$ $\sim 10^{-35} s^{-2}$ is
in agreement with the current status of $\Lambda$
with a small value~ \cite{carmeli02}.\\

{\bf 4.2.~The enigma of the stiff fluid model}\\
It is worthwhile to mention that from the solutions of equation
(9), one may arrive at the solutions obtained by Arbab~
\cite{arbab04} for the dust ($\omega = 0$) case under the {\it
ansatz} $\Lambda = 3\alpha(\dot a/a)^2$. This is possible via the
transformation relation in between $\alpha$ and $\beta$ in the
form $\alpha = \beta/3(\beta - 2)$. By the use of this one can
easily arrive at the equations (10) - (12) which are the same as
those of his equations (8) - (10). However, in the case of stiff
fluid, the expected transformation relation can be obtained from
our equation (15) and his equation (18) which yields the value for
$\beta =3$. It can be seen that this immediately makes the whole
set of Arbab's equation to blow up. This, therefore, suggests that
such transformation is forbidden which is also true in his own
case as this makes the solution set singular. However, exploiting
this transformation relation $\alpha = \beta/3(\beta - 2)$ in the
solution of Majern{\'i}k~\cite{majernik01} it has been shown by
Arbab~ \cite{arbab04} that $\alpha$ can be expressed as having the
form $\kappa/(1+\kappa)$, where $\kappa=\beta/2(\beta -3)$ is a
free parameter. This yields $\alpha =-w_Q$ and $w_Q=p_Q/{\rho}_Q
\quad (-1<w_Q <0)$, being the Quintessence equation of state,
Arbab~ \cite{arbab04} argued that $w_Q$ is nothing but the vacuum
energy parameter. One can easily find out that the $\kappa$ of
Majern{\'i}k~\cite{majernik01} is nothing but our $\gamma$ under
the {\it ansatz} $\Lambda  = 8\pi G\gamma \rho$ and follows the
same relation as shown in equation (15). It suggests that stiff
fluid has a deep connection to {\it dark energy}.

It is also interesting to note that by assuming $n  =
\Omega_{\Lambda}$ (this assumption is not unreasonable since,
$2/3<n<1$) in the equation (27) of the bulk viscous model of Arbab
\cite{arbab97} with variable $G$, we get the same expression for
the age of the universe, viz., $t_0 (= 1/3\Omega_{m0}H_0)$ in our
model. Therefore, it is possible to obtain the same value of $t_0$
for the dust case of Arbab as that obtained in the present stiff
fluid model. Now, it is known that bulk viscosity is associated
with the inflationary universe scenario and after the decoupling
of neutrino in the early universe matter did behave like viscous
fluid~\cite{klimek76}. This bulk viscosity, therefore, is similar
to a variable cosmological term with a repulsive pressure. In the
present investigation, variable $\Lambda$-based stiff fluid model
with constant $G$ and Arbab's viscous model with variable
$\Lambda$ and variable $G$ produce the same result so far as the
present age of the universe is concerned. This again suggests
about the underlying relation between the stiff fluid and the
viscous fluid which has a feature of {\it dark energy}.

In this context we would like to add here that Chakrabarty and
Pradhan \cite{chakra01}, with a time-dependent gravitational
constant $G$, have shown that the cosmological constant $\Lambda$
varies as $t^{-2}$, as in the present stiff fluid case with
constant $G$. So, there seems to exist some connection between
the models with variable $G$ and the present stiff fluid-filled
Zel'dovich universe.\\

{\bf 5.~Concluding Remarks}\\

Although in the present cosmological scenario it is customary to
choose the value of the barotropic index $\omega$ as zero (for
pressureless dust) and sometimes as $1/3$ (for radiation), yet
stiff fluid model which refers to a Zel'dovich universe have been
selected by some authors for various situations such as cold
baryonic universe \cite{zel72}, early hadron era \cite{carr75},
scalar field fluid \cite{madsen92}, for the relativistic situation
prevailing during the early stages of the universe and LRS Bianchi
I cosmological models \cite{chakra01}. There are recent
applications and claims for equation of states in the various
astrophysical realm, e.g. in neutron star RX J1856-3754
\cite{braje02} and  hyperon stars \cite{linares04} which are very
close to the stiff fluid limit (for some more astrophysical as
well as cosmological applications, see  \cite{wesson78} and refs.
therein). In this connection we are interested to mention the work
of Buchert~\cite{buch01} where it has been shown that in the
spatially averaged inhomogeneous cosmologies the averaged
equations show that the averaged scalar curvature must generically
change in the course of structure formation and that an averaged
inhomogeneous perfect fluid, in some cases, act like a free scalar
field source which can be modelled by a 'stiff' fluid.

It is seen that the present age of the universe as calculated in
the standard Friedmann model as well as in some other models
either suffers from low-age problem \cite{over98} or it is as
large as $27.4 \pm 5.6$ Gyr \cite{viswa02} in the `favored' dust
case. In that respect the present stiff fluid model with variable
cosmological constant demands some attention because the age of
the model fits well with the modern accepted limit. But, the
result of this $\Lambda$-stiff fluid model is very much surprising
in the sense that we do not know how this value falls within the
range of the present values of the age of the universe as obtained
in the different $\Lambda$-CDM models. It is already discussed in
the earlier subsection that there exists some kind of underlying
relation in between either stiff fluid and variable $G$ model or
stiff fluid and viscous fluid which has a feature of {\it dark
energy} or with the both. If there is some relation between them
then what is the possible mechanism - that is also not understood.
One possibility may be that the effect of the {\it stiffness} of
the stiff fluid through the process of evolution via radiation to
the present epoch of matter-dominated universe is too weak to
perceive and hence {\it dark energy} in terms of variable
cosmological constant does play its definite role for the
$\Lambda$-stiff fluid model. Thus, here also we are facing a kind
of age problem of the universe which is really intriguing! In this
respect the comment made by Born, on different context, seems
appropriate to quote: ``Whether one or the other of these methods
will lead to the anticipated ``world law''
must be left to future research'' \cite{born62}.\\

{\bf Acknowledgments}\\

One of the authors (SR) would like to express his gratitude to the
authority of IUCAA, Pune, India for providing him Associateship
Programme under which a part of this work was carried out. We are
thankful to Dr. Subharthi Ray for helpful discussions on stiff
fluid equation of state and also to the anonymous referee for his
critical comments and valuable suggestions which helped a lot to
improve the present paper.\\

\end{document}